\documentclass[11pt,fleqn,twoside]{article}
\usepackage{amsfonts,latexsym}
\makeatletter
\newcommand{\prava}{\footnotesize\it
\begin{flushright}
\begin{minipage}{6cm}
Copyright \copyright 1998 by P.G. Est\'evez, E. Conde  and P.R. Gordoa
\end{minipage}
\end{flushright}}

\newcommand{\name}[1]{\begin{flushleft}
                       \LARGE \bf #1
                       \end{flushleft}\vspace{-3mm}}

\newcommand{\Author}[1]{\begin{flushleft}
                       \it #1 \end{flushleft}}

\newcommand{\Adress}[1]{\begin{flushleft}
                       \it #1 \end{flushleft}}

\newcommand{\Date}[1]{\begin{flushleft}
                      \small  \it #1 \end{flushleft}}

\newcommand{\ehkol}{Author \ name}
\newcommand{\ohkol}{Article \ name}
\renewcommand{\@evenhead}{
\hspace*{-3pt}\raisebox{-15pt}[\headheight][0pt]{\vbox{\hbox to \textwidth
{\thepage \hfil \ehkol}\vskip4pt \hrule}}}
\renewcommand{\@oddhead}{
\hspace*{-3pt}\raisebox{-15pt}[\headheight][0pt]{\vbox{\hbox to \textwidth
{\ohkol \hfil \thepage}\vskip4pt\hrule}}}
\renewcommand{\@evenfoot}{}
\renewcommand{\@oddfoot}{}

\newcommand{\be}{\begin{equation}}
\newcommand{\ee}{\end{equation}}
\newcommand{\ba}{\hspace*{-5pt}\begin{array}}
\newcommand{\ea}{\end{array}}

\newcommand{\ds}{\displaystyle}
\makeatother

\begin{document}
\def\theequation{\arabic{section}.\arabic{equation}}
\def \1{\'{\i}}
\language=0

\setcounter{page}{82}

\thispagestyle{empty}

\renewcommand{\ehkol}{P.G. Est\'evez, E. Conde  and P.R. Gordoa}
\renewcommand{\ohkol}{Miura, B\"acklund and Darboux
Transformations}

\begin{flushleft}
\footnotesize \sf
Journal of Nonlinear Mathematical Physics \qquad 1998, V.5, N~1,\ 
\pageref{estevez-fp}--\pageref{estevez-lp}. \hfill
{{\sc Review Article}}
\end{flushleft}

\setcounter{equation}{0}
\setcounter{section}{0}

{\renewcommand{\footnoterule}{}
{\renewcommand{\thefootnote}{}  \footnote{\prava}}

\vspace{-5mm}

\name{Unified approach to Miura, B\"acklund and Darboux
Transformations for Nonlinear \\
Partial Differential Equations}\label{estevez-fp}

\Author{P.G. EST\'EVEZ~$^*$, E. CONDE  and P.R. GORDOA}

\Adress{Area de F\1sica Te\'orica Facultad de F\1sica\\
Universidad de Salamanca, 37008 Salamanca. Spain\\
$^*$~E-mail: pilar@sonia.usal.es}

\Date{Received October 20, 1997}

\begin{abstract}
\noindent
 This paper is an attempt to present and discuss at some length the
Singular Manifold Method.
This Method is based upon the Painlev\'e Property systematically used as a
tool for obtaining clear
cut answers to almost all the questions related with Nonlinear Partial
Dif\/ferential Equations: Lax
pairs, Miura, B\"acklund or Darboux  Transformations as well as
$\tau$-functions, in a unif\/ied way.
Besides to present the basics of the Method we exemplify this approach by
applying it to four
equations in $(1+1)$-dimensions. Two of them are related with the other two
through Miura
transformations that are also derived by using the Singular Manifold Method.
\end{abstract}

\section{Introduction}
\setcounter{equation}{0}

\subsection{Integrability and the Painlev\'e Property}
The beginnings of the study of singularities in the complex plane for
dif\/ferential equations has
always been attributed to Cauchy \cite{estevez:Cauchy}. Cauchy's main idea was to
consider local solutions
on the complex plane and to use methods of analytical prolongation to
obtain global solutions. For
this  procedure to work, a complete knowledge of the singularities of the
equation and its location
in the complex plane is required. In this sense, it is essential to
distinguish between two types
of singularity.

\begin{itemize}
\item {\bf Fixed singularities:}  Singularities determined by the
coef\/f\/icients of the equation
and its location does not therefore depend on initial conditions.

\item {\bf Movable singularities:} Singularities whose location on
the complex plane does indeed depend on  the initial conditions.
\end{itemize}

The truly decisive step towards the elucidation of the relationship
between the analy\-ti\-cal
structure of a  system and its integrability is attributed to the Russian
mathematician {\bf Sof\/ia
Kovalevskaya} \cite{estevez:SK}. Her work focused on the study of the motion of a
rigid solid with a f\/ixed
point from an analysis of the singularities of the solutions \cite{estevez:SK}.
Kovalevskaya's work was
completely new and also addressed to uniquely determine the parameter
values for which the only
movable singularities of the solutions on the complex plane were poles.

Although Kovalevskaya's work was apparently af\/f\/licted by a lack of
followers, in the last decade of
the nineteenth century some mathematicians focused their attention on the
classif\/ication of
ordinary dif\/ferential equations (ODEs) on the basis of the type of
singularity their solutions
were able to exhibit.

It was the French mathematician {\bf Paul Painlev\'e} \cite{estevez:PP} who,
following the ideas of Fuchs,
Kovalevskaya, Picard and others, completely classif\/ied f\/irst order
equations and studied second
order equations. In this last case, he found 50 types of second order
equations whose only movable
singularities were ordinary poles. This special analytical property now
carries his name and
in what follows will be referred to as the {\bf Painlev\'e Property (PP)}
\cite{estevez:WTC}. Of these 50
types of equation, 44 can be integrated in terms of known functions
(Riccati equations, Elliptic
Functions, Linear Equations, etc) and the other six, in spite of having meromorphic solutions, do
not have algebraic integrals that allows one to reduce the equation to
quadratures. Today these
are known as Painlev\'e Transcendents. The main contribution of Paul
Painlev\'e lies in that
he established the basis for a theory that, unlike what had been believed
until then, would
allow one {\it a priori}, by singularity analysis, to decide on the
integrability of an equation
without previously solving it.

Although there is no def\/initive proof of why singularity analysis for an
equation turns out to be
a test of integrability, some circumstances do seem to corroborate this.
For example, it has been
found that Painlev\'e Transcendents often appear in similarity reductions
of equations with solitons
\cite{estevez:AC}, \cite{estevez:KC}. It is also intriguing to note that always a certain
relationship seems
to exist between equations with the PP and Isomonodromy Transformations of
certain linear equations
\cite{estevez:FN}. The validity of Painlev\'e's analysis as a suitable procedure
for detecting integrability
could be related to the combined study of Algebraic Geometry and Abelian
Function theory
\cite{estevez:Tabor}. Jacobi elliptic functions have associated with them a genus 2
Riemann surface
(torus with a hole)  and on this surface all algebraic curves are
meromorphic functions.
Although hyperelliptic functions cannot generally be parametrized in terms
of meromorphic functions,
Jacobi observed that certain combinations of hyperelliptic integrals do
have meromorphic inverses
(this, for example, is the case of the integrals obtained by Kovalevskaya
in the fourth
integrable case for the rigid solid) and these are called Abelian
integrals. Simultaneous study
of these integrals together with the associated Riemann surfaces could be
crucial for
establishing the f\/inal proof of a reliable test of integrability based upon
the ideas f\/irst developed by Paul Painlev\'e and Sof\/ia Kovalevskaya.

}

\subsection{The Ablowitz, Ramani and Segur algorithm}

Between 1955 and 1960  the Korteweg-de Vries equation reappeared in the work
of Fermi-Ulam-Pasta
\cite{estevez:FUP} and in the context of plasma physics. Towards 1965, with Zabusky
and Kruskal \cite{estevez:ZK},
the concept of soliton emerged for the f\/irst time. The Soliton was an
entity describing solitary
wave solutions interacting among themselves without any change in shape
except for a small change in
its phase. With this discovery in mind the Inverse Scattering Technique
(IST) was developed
\cite{estevez:AC}, initially allowing one to solve the KdV equation and then many
integrable equations with
soliton solutions.
The incredible success obtained with the IST technique prompted to
Ablowitz, Ramani and Segur
\cite{estevez:ARS} to develop an algorithm (similar to that used by Kovalevskaya in
the rigid solid problem)
to determine whether an ordinary dif\/ferential equation had the PP. As
mentioned above, an ODE is
said to have the PP if the only movable singularities of its solutions are
poles. This same property
can be stated by saying that all solutions are singlevalued except in the
f\/ixed singularities of
the coef\/f\/icients. The ARS (Ablowitz, Ramani and Segur) algorithm is a
method for determining the
nature of the singularities of the solutions of an ODE on the basis of an
analysis of their local
properties.

Until now we have considered the study of singularities within the context
of systems described
mathematically by ODEs. In view of the results obtained with the ARS
algorithm, it seems natural
to ask ourselves whether some other method for dealing with dif\/ferential
equations in partial
derivatives (PDEs) could in principle be available. In particular, to f\/ind
a new version of the PP
that can be applied directly to the study of integrability for PDEs could
indeed be extremely
useful. One important problem here is that the solutions of a PDE are
functions of at least two
independent variables and their analytical continuation is clearly more
complicated than in the case
of ordinary equations.

The study of similarity reductions for PDEs that can be solved by IST led
Ablowitz, Ramani and
Segur to formulate what has now become know as the {\bf ARS conjecture}: ``Every ordinary
dif\/ferential equation that can be obtained as the similarity reduction of a
PDE solvable by IST has
the PP up to a smooth change of variables".     This conjecture provides a
necessary condition for
checking whether a PDE is integrable or not. Ablowitz, Ramani and Segur
\cite{estevez:ARS} and Mc Leod and
Olver \cite{estevez:McLO} have tested some weak versions of this conjecture. Such
demonstrations are based
on the fact that if a PDE can be completely integrated its solutions can be
evaluated in terms of
the  Gel'fand-Levitan-Marchenko linear integral equation that appears in
the IST.

The ARS conjecture can therefore be understood in the sense that if it is
possible to reduce a
PDE to an ODE that does not have the PP (even after a suitable
transformation of variables) it
may be concluded that the       PDE is not integrable. An algorithmic procedure
has recently been put
forward for determining similarity reductions for PDEs. The essence of the
procedure is the
study of the Lie symmetries. To check that a PDE has the PP using the ARS
conjecture one must f\/ind
all the possible similarity reductions and check that all the resulting
ODEs do have the PP even
though one has to make transformations of variables. In this context the
ARS conjecture is on the
one hand tedious (owing to the huge number of reductions to ODEs shown by
some equations) and on the
other hand becomes less useful as the number of symmetries shown by the
equation decreases.
It is also not clear which transformations of variables are permitted when
checking whether
the corresponding  ODE is of the Painlev\'e type. In particular for
equations that do not have
symmetries the ARS conjecture is quite useless as it is not possible to
obtain similarity
reductions from usual group-theory procedures.

The obvious limitations of this method suggest that it would be interesting
to have available a
direct method (analogue of the ARS algorithm for ODEs) that would allow one
to decide whether
the PDEs under study are integrable. In this way one is extending the
def\/inition of the PP to partial
derivatives using the original idea of Painlev\'e and developing an
algorithmic method to determine
whether the equations have this property (without the requirement of
considering all their similarity
reductions to ODEs) and hence to decide whether they are integrable
or not.

\subsection{The Weiss, Tabor and Carnevale algorithm}

The main dif\/ference between analytical functions of one and several
variables is that the
singularities of the latter are not isolated. If $f  (z_1, \ldots, z_n)$  is
an analytical function
of $n$ complex variables $z_i$ $(i=1,\ldots, n)$, the singularities of
$f$ are in manifolds of
$(2n-2)$-dimensions. These manifolds are determined by conditions of the form
\be
\chi(z_1,\ldots,z_n)=0,
\ee
 where $\chi$ is an analytical function in a
neighborhood of the manifold def\/ined by (1.1).  When this manifold
depends on the initial conditions it is called a {\bf
movable  singularity manifold}. The existence of these singularity
manifolds suggests the need for
introducing the PP concept for PDEs. This leads to a procedure to check
whether the equations
have such a property in a way that makes possible to evaluate the question
of integrability by a
unif\/ied analysis of singularities for both ODEs and PDEs. This was the work
carried out by Weiss,
Tabor and Carnevale (WTC) \cite{estevez:WTC}. According to these authors, we say
that {\bf a PDE has
the Painlev\'e property (PP) if its solutions are singlevalued in a
neighborhood of the manifold of
movable singularities}.

The WTC method also allows one to successfully apply some of the techniques
developed for
integrable systems to systems that are not completely integrable. Certain
restrictions may be
imposed on $\chi$ or on the parameters of the equation such that the
solutions thus obtained
only have poles as movable singularities. In this case, the singularity
manifold $\chi$ is no
longer an arbitrary function and the equation is said to have the
conditional PP \cite{estevez:CT} and hence is partially integrable.

\subsection{The singular manifold method}

Weiss \cite{estevez:W1}, \cite{estevez:W2} introduced the singular manifold method (SMM)
which is an ef\/f\/icient
algorithmic method to f\/ind the typical properties of integrable systems. If
a PDE has the
PP we have seen that its solutions can be expressed as a Laurent series in
the form
\be
u=\sum_{j=0}^{\infty}u_j(z_1,\ldots,z_n)[\chi(z_1,\ldots,z_n)]^{j-a}.
\ee It is
possible in any case to
truncate the expansion series at a certain term in order to obtain
particular solutions of the
equation. If the expansion is truncated at the constant term (understood as
the one that goes with
$\phi^0$ where we use $\phi$ instead of $\chi$ for the manifold in the
truncated expansion), expression (1.2) reduces to:
\be
u=u_0\phi^{-a}+u_1\phi^{1-a}+\cdots +u_a.
\ee

It is interesting to note that some attempts have been made in order to
truncate the Laurent series
at higher orders  \cite{estevez:P}; however here we shall only consider truncation
at the constant
term. Substitution of (1.3) in the corresponding PDE leads to an
overdetermined system of equations
for $\phi$, $u_j$ and their  derivatives. The essential point is that the
singularity manifold
$\phi$ is no longer an arbitrary function but rather -as we now shall see-
it must fulf\/ill certain
equations due to the truncation condition. That's why the truncation of the
Painlev\'e series is
the basis of a method called {\bf Singular Manifold Method} (SMM) that has
been proved to be
extremely successful in studying nonlinear PDEs. Many of the properties of
such equations can be
obtained through the SMM. Let us  summarize some of them.

\begin{itemize}
\item The truncation (1.3) of the Painlev\'e series has itself the
meaning of an
auto-B\"acklund transformation between two solutions of a PDE \cite{estevez:W2},
\cite{estevez:EG96}.

\item  The Lax pair can be obtained through the Singular Manifold equations
\cite{estevez:W1}, \cite{estevez:Musette91}.

\item The relation between the singular manifold method and
nonclassical Lie symmetries of the truncated solutions has been
studied in \cite{estevez:EG95}.

\item   The method of Hirota \cite{estevez:Hir}, \cite{estevez:Hiet}, \cite{estevez:HS}  is
known as a powerful procedure for generating multisoliton solution
for PDEs. It essentially consists in bilinearizing the dif\/ferential
equation by an {\it ansatz} reminiscent of the Painlev\'e
truncated expansion.\\
The WTC method also provides an iterative procedure for generating
solutions \cite{estevez:EG94} from the
Lax pair and from the corresponding auto-B\"acklund transformation, where
the corresponding
singularity manifold $\phi$  is determined in each step and after $n$ steps
the solution can be
expressed in terms of the product $\phi_1$, $\phi_2$, $\ldots$,
$\phi_n$ from which it is then possible
to construct the Hirota $\tau$ function associated with the solution with $n$
solitons. The
relationship between singular manifold and  Hirota's
$\tau$-functions \cite{estevez:EG93}, \cite{estevez:EL}, \cite{estevez:EG97}, \cite{estevez:GL} has also
been clearly
established.

\item The Darboux transformations of a PDE \cite{estevez:MS}, \cite{estevez:A},
\cite{estevez:AN}, \cite{estevez:LU} are also an important procedure to obtain
solutions of PDEs. The connection between SMM and Darboux
transformations has been explained in dif\/ferent references  \cite{estevez:EL},
\cite{estevez:EG97}.
\end{itemize}

\subsection{Plan of the paper}

After the previous glimpses of evidence in regard to the relationship
between the PP and the
integrability conditions various directions in the search for integrable
PDE become evident. On
the one hand the Painlev\'e test does identify integrable systems and on
the other hand the Singular
Manifold Method appears as a systematic technique for f\/inding B\"acklund
and Darboux transformations,
Lax pairs, Soliton Solutions etc. Our point of view -as we shall show
henceforth- is that the PP
should be used not only as a test of integrability but also as a fruitful
source of information of
practically all the important features of the Non Linear Partial
Dif\/ferential Equations. This paper
tries to go an step further in this direction, adding the Miura transformations to the above
mentioned properties arising from the SMM.

Sections 2 and 4 deal with the application of the SMM to two equations in
$(1+1)$-dimensions as
the AKNS (Ablowitz-Kaup-Newell-Segur) \cite{estevez:AKNS} and non local Boussinesq
equation NLBq
\cite{estevez:LW}, \cite{estevez:WL} equation. The Lax pair, Darboux transformations and
Solitonic Solutions are
thus fully obtained in these cases. This also shows that our analysis
becomes not only a
conceptual piece of information but also an algorithmic tool that can be
systematically used.

Sections 3 and 5 are devoted to the study of the ShG (sinh-Gordon)
\cite{estevez:AC} and KS (Kaup system)
\cite{estevez:Kaup} systems. The Miura transformations between these equations and
AKNS and NLBq
respectively are obtained by using the SMM. B\"acklund transformations for
AKNS and NLBq are
derived. The two component induced Lax pairs for ShG  and KS are identif\/ied
by the same procedure.

Section 6 is one of conclusions. Some lengthy and/or auxiliary calculations
are relegated to Appendices A to E.

\section{The AKNS equation in (1+1)-dimensions}
\setcounter{equation}{0}

The well known \cite{estevez:AKNS} AKNS equation  in 1+1,
\be
 0=M_{yxxx}+4M_yM_{xx}+8M_xM_{xy}\ee
is a nice and easy example to start to show how the method works.

The PP for this equation means that all solutions of (2.1) can be written
as a series  of the form
(see  \cite{estevez:WTC}):
\be
M=\sum_{j=0}^{\infty}M_j\chi^{j-a},
\ee
where $a$ is the leading index and $\chi$ is an arbitrary function of $x$
and $t$,
depending on the initial data, that is usually called {\bf singularity
manifold}. $M_j$ are
analytical functions of $t$ and $x$ in the neighborhood of $\chi=0$. It is
a trivial exercise to
substitute (2.2) into (2.1) and to check that if the leading index is
$a=1$, the series (2.2) satisf\/ies (2.1) for any functional form of $\chi$.
It could be
said that (2.1) has the PP.

\subsection{Truncated expansion. B\"acklund transformations}

The SMM is based upon the above def\/ined PP. It requires the truncation of
(2.2) at the constant
level
$j=a$. It means, for our equation (2.1), that the truncated solutions
should~be:
\be
M'= M +{\phi_x\over\phi},
\ee
where we have called $ M=M_1$ and $M'$ to the truncated solution. We have
used here $\phi$
for the singularity manifold, instead of $\chi$, to emphasize  that the
truncation implies that the
singularity manifold is no longer an arbitrary function but a function that
is ``singularized" by
the fact that it  should satisfy some def\/inite equations as we will see
later. We  shall be  calling
this manifold $\phi$ {\bf singular manifold} \cite{estevez:W1}, \cite{estevez:W2} and the
method based on the
truncation of the  Painlev\'e series is the so called Singular Manifold Method.

The substitution of the truncated expansion (2.3) into the equation (2.1)
provides the following results (see Appendix A):

\medskip

$\bullet$ $ M$ as well as $M'$ should be solutions of (2.1). It means that
(2.3) could be consided as an auto-B\"acklund transformation between
two solutions $M'$ and $ M$ of the same equation.

\medskip

$\bullet$ The solution $ M$ can be written in terms of the singular
manifold in the following way:
\be
M_x=-\left({1\over 4}\right)\left(v_x+{v^2\over
2}+2\lambda\right),
\ee
\be
 M_y={1\over 2}(-v_y+2\lambda q).
\ee
The notation that we have used is \cite{estevez:EG97}:
\be
v={\phi_{xx}\over \phi_x},
\ee
\be q={\phi_{y}\over \phi_x},
\ee
and $\lambda$ is an arbitrary constant that, as we will see, plays the role
of the spectral parameter.

\medskip

$\bullet$ {\bf The singular manifold equations}. The equations that the
truncation procedure implies for $\phi$ are:
\be
s_y=4\lambda q_x,
\ee
where $s$ is the schwartzian derivative def\/ined as:
\be
s=v_x-{v^2\over 2}.
\ee
Furthermore the compatibility condition $\phi_{xxt}=\phi_{txx}$ between the
def\/initions (2.6) and (2.7) requires:
\be
v_y=(q_x+qv)_x\Longrightarrow s_y=q_{xxx}+2sq_x+qs_x.
\ee
It is not dif\/f\/icult to prove that the singular manifold equations are
nothing but the AKNS system once again. In fact with the change of
variables:
\be
 s=4p_x+2\lambda,
\ee
\be
q={p_y\over \lambda}.
\ee
(2.8) is trivially fulf\/illed and (2.10) is written as:
\be
 0=p_{yxxx}+4p_yp_{xx}+8p_xp_{xy}
\ee
that is obviously the AKNS system.

\subsection{Lax pairs}

As we have seen above, the singular manifold equations, written in terms of
$v$ and $w$ are:
\be
v_{xy}-vv_y=4\lambda q_x,
\ee
\be
v_y=(q_x+qv)_x
\ee
that can be considered as a new system of nonlinear equations. If we apply
the Painlev\'e analysis
to this system, the leading terms are (using $\psi$ for the singularity
manifold).
\[
v\sim v_0\psi^a, \qquad q\sim q_0\psi^b.
\]
The substitution in (2.14-15) provides:
\[
a=-1,\qquad b=-2,\qquad v_0=2\psi_x,
\qquad q_0=-{1\over \lambda}\psi_x\psi_y.
\]
These leading terms provide the key for the linearization of the truncated
solutions (2.3-4).
Actually if we substitute  $v$ for its dominant term.
\be
v=2{{\psi}_x\over {\psi}}\Longrightarrow \phi_x={\psi}^{2}.
\ee
In such a case (2.4) is:
\be
0=\psi_{xx}+(2 M_x+\lambda)\psi
\ee
and from (2.16), (2.5) and (2.10) we obtain:
\[
2{\psi_y\over \psi}=q_x+qv={1\over
2\lambda}(2M_{xy}+v_{xy}+2vM_y+vv_y)
\]
or
\be
0=2\lambda \psi_y+ M_{xy}\psi-2 M_y\psi_x.
\ee
(2.17) and  (2.18) are precisely the Lax pair for AKNS. To summarize it is
possible to say that
{\bf the Lax pair is nothing but the singular manifold equations in which
the eigenfuctions are
directly obtained from the singular manifold} through (2.16).

\subsection{Darboux transformations}

Following an idea of Konopelchenko and Stramp \cite{estevez:KS}, we can consider
the Lax pair itself as a
pair of coupled nonlinear equations between $M$ and $\psi$. Let us now
explain how to proceed.

As far as $M'$ is also a solution of (2.1) an associated singular manifold
$\phi_2'$ linked to an
spectral parameter $\lambda_2$ can be def\/ined just by def\/ining
\be
\phi'_{2x}=\psi_2^{'2}
\ee
a Lax pair for $M'$ can be written as
\be
0=\psi'_{2xx}+(2 M'_x+\lambda_2)\psi'_2,
\ee
\be
0=2\lambda_2 \psi'_{2y}+ M'_{xy}\psi'_2-2 M'_y\psi'_{2x},
\ee
where the notation means that $\psi'_2$ is an eigenfunction for $M'$ with
eigenvalue $\lambda_2$.
If we call $\phi_1$ and $\phi_2$ two singular manifolds for $M$ attached to
spectral parameters
$\lambda _1$ and $\lambda_2$ respectively the corresponding eigenfunctions
are def\/ined as
\be
\phi_{1x}=\psi_1^{2},
\ee
\be
\phi_{2x}=\psi_2^{2},
\ee
and the Lax pairs take the form
\be
0=\psi_{1xx}+(2 M_x+\lambda_1)\psi_1,
\ee
\be
0=2\lambda_1 \psi_{1y}+ M_{xy}\psi_1-2 M_y\psi_{1x},
\ee
\be
0=\psi_{2xx}+(2 M_x+\lambda_2)\psi_2,
\ee
\be 0=2\lambda_2 \psi_{2y}+ M_{xy}\psi_2-2 M_y\psi_{2x}.
\ee

If we use the singular manifold $\phi_1$ to construct the truncated
Painlev\'e expansion
\be
M'=M+{\phi_{1x}\over \phi_1}
\ee
and we then look at (2.20-21) as a system of nonlinear coupled equations  a
similar expansion should be performed for $\psi'_2$. That is to say:
\be
\psi'_2=\psi_2 +{\Theta\over \psi_1}.
\ee

The substitution of the truncated expansions (2.28-29) in (2.20-21)
provides the functional form for
$\Theta$. The result is
\be
 \Theta =-\psi_1\Omega(\psi_1,\psi_2),
\ee
where
\be
\Omega(\psi_1,\psi_2)=\left({1\over\lambda_1-\lambda_2}\right)(\psi_1\psi_{2
x}-\psi_2\psi_{1x}).
\ee

The expansions (2.28-29) can be considered as transformations that leave
invariant the Lax pair
(2.18-19). In this sense they are Darboux transformations. It should be
pointed out that we use the
singular manifold $\phi_1$ to actually realize the transformation (2.28)
but not the eigenfunction
$\psi_1$ as it is usual \cite{estevez:MS} in the Darboux transformations.
Nevertheless eigenfunctions and
Singular Manifolds are trivially related through (2.22) and therefore: {\bf
With two eigenfunctions
$\psi_1$ and $\psi_2$ for
$M$, we can construct an eigenfunction $\psi'_2$ for the iterated solution
$M'$}. That is
why we call them Darboux transformations.

\subsection{Hirota's function}

Furthermore, (2.19) is a nonlinear equation that relates $\phi'_2$ and
$\psi'_2$. It means that
the singular manifold $\phi'_2$ itself could also be expanded in terms of
$\phi_1$
\be
\phi'_2=\phi_2+{\Delta\over \phi_1}
\ee
and by substituting this expansion in (2.19) we obtain:
\be
\Delta=-[\Omega(\psi_1,\psi_2)]^2.
\ee
The procedure described above could be easily iterated. The singular
manifold $\phi'_2$ for $M'$
can be used to construct a new solution
\be
M''=M'+{\phi'_{2x}\over \phi'_2}
\ee
that combined with (2.28) can be written as:
\be
M''=M+{\tau_{12x}\over \tau_{12}},
\ee
where
\be
\tau_{12}=\phi'_2\phi_1
\ee
and by using (2.32) and (2.33)
\be
\tau_{12}= \phi_2\phi_1-[\Omega(\psi_1,\psi_2)]^2.
\ee
It is an interesting point to note that the function $\tau_{12}$ for the
second iteration is
not a Singular Manifold  but it  can be constructed from two Singular
Manifolds of the f\/irst
iteration. The SMM provides algorithmically, but sometimes considered just a
clever ansantz \cite{estevez:GL}, \cite{estevez:Hiet},
the Hirota's bilinear method \cite{estevez:Hir}. It also
provides the way
to construct solutions for the $\tau$-function, as we will see in the
next section.

\subsection{Solitonic solutions}

The easiest nontrivial solutions can be obtained from the seminal solution
\be
M=a_0y.
\ee
For this solution, exponential solutions of (2.24-27) are
\be
\psi_i=\exp\left({k_ix-{a_0\over k_i}y}\right),
\ee
where
\be
\lambda_i=-k_i^2
\ee
and the corresponding manifolds are
\be
\phi_i={1\over 2k_i}(\alpha_i+\psi_i^2),
\ee
where $\alpha_i$ are arbitrary constants. The equation (2.31) provides:
\be
\Omega(\psi_1,\psi_2)={1\over k_1+k_2}\psi_1\psi_2
\ee
and (2.37) also yields
\be
\tau_{12} = {1\over
4k_1k_2}(\alpha_1+\psi_1^2)(\alpha_2+\psi_2^2)-{\psi_1^2\psi_2^2\over
(k_1+k_2)^2}.
\ee
We can write the f\/irst and second iteration as:
\be
M'=a_0y+{\phi_{1x}\over \phi_1},
\ee
\be
M''=a_0y+{\tau_{12x}\over \tau_{12}},
\ee
where
\be
\phi_1= {\alpha_1\over 2k_1}(1+F_1),
\ee
\be
\tau_{12}={1\alpha_1\alpha_2\over
4k_1k_2}\left\{1+F_1+F_2+A_{12}F_1F_2\right\},
\ee
\[
\alpha_i=\exp\left({2k_ix_{0i}}\right),
\]
\be
F_i=\exp\left({2k_i\left(x-{a_0\over k_i^2}y-x_{0i}\right)}\right),
\ee
\be
A_{12}=\left({k_1-k_2\over k_1+k_2}\right)^2.
\ee
(2.44) corresponds to the one-soliton solution  (see Fig.~1) and
(2.45)  to the interaction of two solitons  (see Fig.~2).

\vspace*{-2mm}

\section{The sinh-Gordon equation. \\
Miura Transformation to AKNS}

\setcounter{equation}{0}
We shall be analyzing in this section how the SMM and the Singular Manifold
equations are able to provide also information about Miura
transformations between nonlinear PDE's. In particular, this section
is devoted to obtain the Miura map between AKNS and the sinh-Gordon
equation.

The sinh-Gordon equation \cite{estevez:AC}
\be
U_{xy}=\sinh 2U
\ee
can be written as the system
\be
0= u_{xy}+2u\eta_y,
\ee
\be
0=\eta_x+u^2
\ee
through the change
\be
u=U_x,
\ee
\be
\eta_y=-\cosh 2U.
\ee
This system has the same problem as the sine-Gordon equation
\cite{estevez:CM}. It has two Painlev\'e branches. In fact if we write the
solutions of (3.2-3.3) as a Painlev\'e series
\[
\ba{l}
\ds u=\sum\limits _{j=0}^{\infty}u_j\chi^{j-a},\\[3mm]
\ds \eta=\sum\limits _{j=0}^{\infty}\eta_j\chi^{j-b}.
\ea
\]
The leading indexes are $a=b=1$ but the dominant terms are:
\be
u_0=\pm \chi_x,
\ee
\be
\eta_0=\chi_x.
\ee
The $\pm\,\mbox{sign}$ of $u_0$ means that there are two
possibilities for the expansion. This may seem at
f\/irst a problem when one attempts to apply the SMM since if we choose
a def\/inite sign in the expansion we will be  loosing
information about the equation. This problem has
been discussed during the last years (see \cite{estevez:EG93}, \cite{estevez:EG97}, \cite{estevez:CM},
\cite{estevez:CMP}). In these papers  a modif\/ication of the SMM appears to be
necessary. We need in fact
to deal with two Singular Manifolds altogether. Obviously  this fact
represents a non trivial complication for the calculations.

We present here the easiest form to work with these two Singular Manifolds.
The idea is the
following: The two branches (3.6-7) suggest the following set of changes
for the functions:
\be
u=m-\hat m,
\ee
\be
\eta=m+\hat m
\ee
in such a way that $m$ and $\hat m$ should have only a Painlev\'e branch. It
is now necessary to look for the equations that $m$ and $\hat m$ should
satisfy. For this
purpose  we introduce in (3.2-3)  the change (3.8-9). Adding and
subtracting the result we obtain:
\be
m_{xy}+2(m-\hat m)m_y=0,
\ee
\be
\hat m_{xy}-2(m-\hat m)\hat m_y=0.
\ee
From these equations we also obtain the additional information that
we discuss in the next Subsections.

\subsection{Miura Transformation}

The $\hat m$ can be obtained from (3.10) and substituted in (3.11). The
result is that $m$
obeys:
\be
0=2m_ym_{xxy}+8m_xm_y^2-m_{xy}^2
\ee
that is the integrated version of the AKNS equation
\be
0=m_{yxxx}+4m_ym_{xx}+8m_xm_{xy}.
\ee
In a similar form, we can obtain $m$ from (3.11) and the substitution
in (3.10) is
\be
0=2\hat m_y\hat m_{xxy}+8\hat m_x\hat m_y^2-\hat m_{xy}^2
\ee
that it is again the integration of the AKNS equation
\be
0=\hat m_{yxxx}+4\hat m_y\hat m_{xx}+8\hat m_x\hat m_{xy}.
\ee
As a consequence of what has just been said the change of functions (3.8-9)
can be inverted to yield
\be
2m_x=u_x+\eta_x=u_x-u^2,
\ee
\be
2\hat m_x=\eta_x-u_x=-u_x-u^2
\ee
which represents the Miura transformations between the sinh-Gordon system
(3.2-3) and the AKNS equations (3.13) and (3.15).

\subsection{B\"acklund  Transformations}

As an additional result we obtain that the solutions $(u,\eta)$ of
sinh-Gordon can be constructed by using two solutions $m$ and $\hat
m$ of AKNS.
Nevertheless these solutions
are not independent. They are related by (3.10-11) that can be written as:
\be
\hat m=m+{1\over 2}{m_{xy}\over m_y}, \qquad\qquad  m=\hat m+{1\over
2}{\hat m_{xy}\over\hat m_y},
\ee
\be
(m_y\hat m_y)_x=0.
\ee
These equation (3.18-19) can easily be recognized as B\"acklund
transformations between the two solutions $m$ and $\hat m$ of AKNS.

\subsection{Singular manifold method with two manifolds}

\noindent
From the above discussion it is easy to understand why it seems reasonable
to talk about two
Singular Manifolds, one for the expansion of $m$ and the other for the
expansion on $\hat m$. Let us
call $\phi$  the singular manifold for $m$ and $\hat \phi$ the singular
manifold for $\hat m$. The
truncated expansions are
\be
m'=m+{\phi_x\over \phi},
\ee
\be
\hat m'=\hat m+{\hat \phi_x\over \hat \phi},
\ee
and the corresponding expansions for $u$ and $\eta$ are:
\be
u'=u+{\phi_x\over \phi}-{\hat \phi_x\over \hat \phi},
\ee
\be
\eta'=\eta+{\phi_x\over \phi} +{\hat \phi_x\over \hat \phi}.
\ee
However $\phi$ and $\hat \phi$ are not independent because $m$ and
$\hat m$ are
related by the B\"acklund transformation (3.18-19). This is reminiscent of
the requirement
that $u$ and $\eta$ satisfy (3.2-3). In fact substituting (3.22-23) in
(3.3) we obtain
the coupling condition between $\phi$ and $\hat \phi$  (see Appendix B)
\be
{\phi_x\over \phi} {\hat \phi_x\over \hat \phi}=A{\phi_x\over \phi}
+\hat A{\hat \phi_x\over \hat \phi},
\ee
where
\be
A={v\over 2}+u,
\ee
\be
\hat A={\hat v\over 2}-u.
\ee
The notation is the one def\/ined in (2.6).

\subsection{Lax pair for sinh-Gordon}

The derivative of (3.24) with respect to $x$ (see Appendix B) provides:
\be
A_x=A(\hat v-A-\hat A)={\hat v-v\over 2},
\ee
\be
\hat A_x=\hat A( v-A-\hat A)={ v-\hat v\over 2}.
\ee
We should remember at this point that $\phi$ and $\hat \phi$ are Singular
Manifolds for AKNS and that the change
\be
\phi_x=\psi^2, \qquad\qquad \hat\phi_x=\hat\psi^2
\ee
relates the Singular Manifolds with the eigenfunctions of the Lax pair
(2.17-18) of AKNS.
By combining (3.29) with (3.27-28) we can integrate out the variable $x$,
and f\/inally obtain
\be
A=a{\hat \psi\over \psi},
\ee
\be
\hat A=\hat a{ \psi\over \hat\psi},
\ee
where $a$ and $\hat a$ are constants. (3.25-26) can be now written as:
\be
 \psi_x=a\hat \psi-u\psi,
\ee
\be
\hat \psi_x=\hat a \psi+u\hat\psi,
\ee
where $\psi$ and $\hat \psi$ are solutions of the Lax pair of AKNS. That is
equivalent to
\be
0=\psi_{xx}+(2 m_x+\lambda)\psi,
\ee
\be
0=2\lambda \psi_y+ m_{xy}\psi-2 m_y\psi_x,
\ee
\be
0=\hat\psi_{xx}+(2 \hat m_x+\hat \lambda)\hat \psi,
\ee
\be
0=2\hat \lambda \hat\psi_y+ \hat m_{xy}\hat \psi-2 \hat
m_y\hat\psi_x.
\ee
The compatibility between (3.32-33) and (3.34), (3.36) requires
\be
\lambda=\hat \lambda=-a\hat a.
\ee
The $y$ component of the Lax pair is given by (3.35), (3.37) and can be
written as
\be
2\hat a\psi_y=-(u_y+\eta_y)\hat \psi,
\ee
\be
2 a\hat \psi_y=(u_y-\eta_y) \psi.
\ee

Therefore, the Lax pair for sinh-Gordon can be written in its usual
matrix form as:
\be
\left(\begin{array}{c}\psi\\ \hat \psi\end{array}
\right)_x
=
\left(
\begin{array}{cc} -u &a\\ \hat a  &u
\end{array}
\right)\left(\begin{array}{c}\psi\\ \hat \psi\end{array}
\right),
\ee
 \be
\left(\begin{array}{c}2\hat a\psi\\2 a\hat \psi\end{array}
\right)_y
=
\left(
\begin{array}{cc} 0 &u_y+\eta_y\\u_y-\eta_y &0
\end{array}
\right)\left(\begin{array}{c}\psi\\ \hat \psi\end{array}
\right).
\ee

\noindent{\bf Conclusion}: The method that we derived in Section 2 to
obtain Darboux
transformations and solutions for AKNS can be directly applied to
sinh-Gordon. The
construction of solution of sinh-Gordon can be done through (3.8-9)  by
using two
solutions of AKNS related by the B\"acklund transformation (3.18). From the
point of view
of the Singular Manifold this implies that $\phi$ and $\hat \phi$ are
related by the
coupling condition (3.24). Using (3.29-31) this condition can be written in
a much easier form as:
\be
a\hat\phi+\hat a\phi=\psi\hat\psi,
\ee
\be
\hat \lambda=\lambda=-a\hat a.
\ee

We believe that the splitting (3.8-9) is the key to solve the long standing
problem that concerns
to the application of the SMM to equations with two branches. In the next
sections we will return
to the same topic in a dif\/ferent but not unrelated context.

\section {Non local Boussinesq equation}
\setcounter{equation}{0}

The following system of equations
\be
N_x=M_t,
\ee
\be
M_xN_t=M_xM_{xxx}+2M_x^3+M_t^2-M_{xx}^2
\ee
or equivalently
\be
M_x^2(M_{tt}-M_{xxxx
})=4M_x^3M_{xx}+2M_x(M_tM_{tx}-M_{xx}M_{xxx})-M_{xx}(M_t^2-M_{xx}^2)
\ee
has been considered in \cite{estevez:LW}, \cite{estevez:WL} as related with the Kaup
(sometimes called Classical
Bousinesq) system through a Miura transformation. The Kaup system is a good
example of system
with two branches \cite{estevez:Kaup}, \cite{estevez:EG93}, \cite{estevez:CMP}.  Nevertheless
(4.1-2) has only one branch.
That is why we are interested in this section in studying the relationship
of both equations
from the point of view of the SMM and to relate the corresponding  results
with the Kaup system in Section 5.

\subsection{Truncated expansion. B\"acklund transformations}

The leading terms of (4.1-2) are
\[
M\sim \chi_x\chi^{-1},
\]
\[
N \sim \chi_t\chi^{-1}
\]
that suggests the truncated expansion
\be
M'= M +{\phi_x\over\phi}\longrightarrow  N'= N +{\phi_t\over\phi}.
\ee
From this equations we obtain the following set of results (see Appendix C):


\medskip

$\bullet$  (4.3) could be consided as an auto-B\"acklund transformation
between two solutions $M'$ and $ M$ of the same equation.

\medskip

{\samepage
$\bullet$ The solution $M$ can be written in terms of the Singular
Manifold as:
\be
 M_x=\left({1\over 4}\right)\left((w+2\lambda)^2-v^2\right),
\ee
\be
M_t={1\over 2}\left\{(w+2\lambda)v_x-vw_x+
(w+\lambda)\left[(w+2\lambda)^2-v^2\right]\right\},
\ee
where \cite{estevez:EG97}:
\be
v={\phi_{xx}\over \phi_x},
\ee
\be
w={\phi_{t}\over \phi_x},
\ee
and $\lambda$ is an arbitrary constant.}

\medskip

$\bullet$ {\bf The singular manifold equations} can be written as the
system of PDEs:
\be
v_t=(w_x+wv)_x
\ee
\be
w_t=\left (v_x-{v^2\over 2}+{3\over 2}(w+2\lambda)^2-2\lambda
(w+2\lambda)\right)_x.
\ee
Notice that the parameter $\lambda$ can be removed from these equations
through the galilean transformation
\[
\bar w\rightarrow (w+2\lambda),
\]
\[
\bar x\rightarrow x-2\lambda t
\]
that transforms (4.9-10) into
\be
v_t=(\bar w_{\bar x}+\bar wv)_{\bar x},
\ee
\be
\bar w_t=\left (v_{\bar x}-{v^2\over 2}+{3\over 2}\bar
w^2\right)_{\bar x}.
\ee
This system is equivalent to the Kaup or classical Boussinesq system. In
fact it could be written as a single equation if we set
\[
\bar w =p_{\bar x}.
\]
One  can now remove $v$ from (4.11-12) and the result is a sort of modif\/ied
Boussinesq equation
\be
p_{tt}-p_{\bar x\bar x\bar x\bar x}-4p_{\bar x}p_{\bar x t}+6p_{\bar
x}^2p_{\bar x \bar x}-2p_tp_{\bar x\bar x}=0.
\ee
Unlike the AKNS case in which the Singular Manifold equations were also
AKNS, for NLBq the Singular Manifold equations (4.11-12) are not the
same system but rather they become the Kaup system (KS). As we will
see in the next section this means that both systems NLBq and KS are
related by a Miura transformation.

\subsection{Lax pairs}

The question of the linearization of the singular manifold equations is now
a little bit more
complicated that it was for AKNS. If we look for the dominant terms of (4.11-12)
\[
v\sim v_0\chi^a,
\]
\[
\bar w\sim \bar w_0\chi^b.
\]
The result is:
\[
a=-1,\qquad b=-1,\qquad v_0=\chi_{\bar x},\qquad \bar
w_0=\pm\chi_{\bar x},
\]
the $\pm\,\mbox{sign}$ conf\/irms the well known fact that the Kaup system has two
Painlev\'e branches
\cite{estevez:EG93}, \cite{estevez:CMP}. In the previous section we have explained that for
those types of systems it is necessary the introduction of two
Singular Manifolds. From that point of view, the
truncation ansatz for $v$ and $\bar w$ is now:
\be
v={{\psi^+}_{\bar x}\over {\psi^+}}+{{\psi^-}_{\bar x}\over
{\psi^-}}\Longrightarrow
\phi_{\bar x}={\psi}^+ {\psi}^-,
\ee
\be
\bar w={{\psi^+}_{\bar x}\over {\psi^+}}-{{\psi^-}_{\bar x}\over
{\psi^-}}
\ee
or
\be
2{{\psi^+}_{x}\over {\psi^+}}=v+w+2\lambda,
\ee
\be
2{{\psi^-}_{x}\over {\psi^-}}=v-w-2\lambda.
\ee

With this ansatz the  expressions (4.5-6)  for  the truncated solutions
(see Appendix D) can be linearized as:
\be
0=2M_x(\psi^+_{xx}+M_x\psi^+)-(M_t+M_{xx}+2\lambda M_x)\psi^+_{x},
\ee
\be
0=\psi^+_t-\psi^+_{xx}+2\lambda \psi^+_x-2M_x\psi^+,
\ee
\be
0=2M_x(\psi^-_{xx}+M_x\psi^-)+(M_t-M_{xx}+2\lambda M_x)\psi^-_{x},
\ee
\be
0=\psi^-_t+\psi^-_{xx}+2\lambda \psi^-_x+2M_x\psi^- .
\ee

To summarize, the existence of two branches in the Singular Manifold
equations implies the
existence of two classes of eigenfunctions of two dif\/ferent Lax pairs
(4.18-19) and (4.20-21).
This result ref\/lects the fact that the equation is invariant under the
discrete symmetry
\[
\ba{rcl}
x&\longrightarrow &-x, \\
t&\longrightarrow &-t,\\
M&\longrightarrow & -M
\ea
\]
that transforms the f\/irst of the Lax pairs into the second one.

\subsection{Darboux transformations }

The generation of the Darboux transformations can be done in a similar way
as we did in Section 2.
If we write the Lax pairs for the iterated solution $M'$
\be
0=2M'_x(\psi^{'+}_{2xx}+M'_x\psi^{'+}_2)-(M'_t+M'_{xx}+2\lambda_2
M'_x)\psi^{'+}_{2x},
\ee
\be
0=\psi^{'+}_{2t}-\psi^{'+}_{2xx}+2\lambda_2
\psi^{'+}_{2x}-2M'_x\psi^{'+}_2,
\ee
\be
0=2M'_x(\psi^{'-}_{2xx}+M'_x\psi^{'-}_2)+(M'_t-M'_{xx}+2\lambda_2
M'_x)\psi^{'-}_{2x},
\ee
\be
0=\psi^{'-}_{2t}+\psi^{'-}_{2xx}+2\lambda_2
\psi^{'-}_{2x}+2M'_x\psi^{'-}_2,
\ee
where $\psi_2^{'+}, \psi_2^{'-}$ could be related to a singular manifold
$\phi_2'$ in the form
\be
\phi'_2=\psi_2^{'+}\psi_2^{'-}.
\ee
The consideration of (4.22-25) as coupled nonlinear equations allows us to
write truncated
expansions for $M',\psi_2^{'+}, \psi_2^{'-},\phi'_2$.
\be
M'=M+{\phi_{1x}\over \phi_1},
\ee
\be
\psi_{2}^{'+}=\psi^+_2 +{\Theta^+\over \phi_1},
\ee
\be
\psi_2^{'-}=\psi^-_2 +{\Theta^-\over \phi_1},
\ee
\be
\phi_2^{'}=\phi_2 +{\Delta\over \phi_1},
\ee
where
\be
0=2M_x(\psi^{+}_{2xx}+M_x\psi^{+}_2)-(M_t+M_{xx}+2\lambda_2
M_x)\psi^{+}_{2x},
\ee
\be
0=\psi^{+}_{2t}-\psi^{+}_{2xx}+2\lambda_2
\psi^{+}_{2x}-2M_x\psi^{+}_2,
\ee
\be
0=2M_x(\psi^{-}_{2xx}+M_x\psi^{-}_2)+(M_t-M_{xx}+2\lambda_2
M_x)\psi^{-}_{2x},
\ee
\be
0=\psi^{-}_{2t}+\psi^{-}_{2xx}+2\lambda_2
\psi^{-}_{2x}+2M_x\psi^{-}_2,
\ee
\be
\phi_{2x}=\psi_2^{+}\psi_2^{-},
\ee
\be
0=2M_x(\psi^{+}_{1xx}+M_x\psi^{+}_1)-(M_t+M_{xx}+2\lambda_1
M_x)\psi^{+}_{1x},
\ee
\be
0=\psi^{+}_{1t}-\psi^{+}_{1xx}+2\lambda_1
\psi^{+}_{1x}-2M_x\psi^{+}_1,
\ee
\be
0=2M_x(\psi^{-}_{1xx}+M_x\psi^{-}_1)+(M_t-M_{xx}+2\lambda_1
M_x)\psi^{-}_{1x},
\ee
\be
0=\psi^{-}_{1t}+\psi^{-}_{1xx}+2\lambda_1
\psi^{-}_{1x}+2M_x\psi^{-}_1,
\ee
\be
\phi_{1x}=\psi_1^{+}\psi_1^{-}.
\ee

By substituting the truncation (4.27-30) in (4.22-26) it is possible
to obtain
\be
\Theta^+ =-\psi_1^+\Omega^+,
\ee
\be
\Theta^- =-\psi_1^-\Omega^-,
\ee
\be
\Delta=-\Omega^+\Omega^-,
\ee
where
\be
\Omega^+=\left({1\over \lambda_2-\lambda_1}\right)
{\psi_1^-\over\psi_{1x}^+}(\psi^+_1\psi^+_{2x}-\psi^+_2\psi^+_{1x}),
\ee
\be
\Omega^-=\left({1\over \lambda_2-\lambda_1}\right)
{\psi_2^-\over\psi_{2x}^+}(\psi^+_1\psi^+_{2x}-\psi^+_2\psi^+_{1x}).
\ee
(4.27-30) together with (4.41-45)  def\/ine Darboux transformations for NLBq.

\subsection{Hirota's function}

The generation of a new iteration can be done by using $\phi'_2$ as the
Singular Manifold for $M'$ in order to construct
\be
M''=M'+{\phi'_{2x}\over \phi'_2}=M+{\tau_{12x}\over \tau_{12}},
\ee
where
\be
\tau_{12}=\phi'_2\phi_1
\ee
and by using (4.30) and (4.43)
\be
\tau_{12}= \phi_2\phi_1- \Omega^+\Omega^-.
\ee

\subsection{Solitonic Solutions}

The easiest nontrivial solutions can be obtained from the seminal solution
\be
M=a_0x.
\ee
For this solution, exponential solutions of (4.22-25) are
\be
\psi_i^+=\exp\{{a_i(x-a_it)}\},\qquad\qquad \psi_i^-=
\exp\left\{-{a_0\over a_i}\left(x-{a_0\over a_i}t\right)\right\},
\ee
where $a_i$ are related
with the spectral
parameter in the form
\be
\lambda_i=a_i+{a_0\over a_i}
\ee
and the corresponding singular manifolds are
\be
\phi_i={a_i\over a_i^2-a_0}\,(\alpha_i+\psi_i^+\psi_i^-),
\ee
where $\alpha_i$ are arbitrary constants.
(4.44-45) gives:
\be
\Omega^+={a_2\over a_2a_1-a_0}\psi_1^-\psi_2^+,
\qquad\qquad
\Omega^-={a_1\over a_2a_1-a_0}\psi_1^+\psi_2^-,
 \ee
and (4.48) is:
\be
\tau_{12} = {a_1a_2\over (a_1^2-a_0)
(a_2^2-a_0)}\,(\alpha_1+\psi_1^+\psi_1^-)(\alpha_2+\psi_2^+\psi_2^-)-{a_1a_2
\over
(a_2a_1-a_0)^2}\psi_1^-\psi_2^+\psi_1^+\psi_2^-.
\ee
 We can write the f\/irst (Fig.~3) and second iteration (Fig.~4) as:
\be
M'=a_0x+{\phi_{1x}\over \phi_1},
\ee
\be
M''=a_0x+{\tau_{12x}\over \tau_{12}},
\ee
where
\be
\phi_1= {\alpha_1 a_1\over a_1^2-a_0}(1+F_1),
\ee
\be
\tau_{12}={a_1a_2\alpha_1\alpha_2\over
(a_1^2-a_0)(a_2^2-a_0)}\left\{1+F_1+F_2+A_{12}F_1F_2\right\},
\ee
\[
\alpha_i=\exp\left\{\left(a_i-{a_0\over a_i}\right)x_{0i}\right\},
\]
\be
F_i={\psi_i^+\psi_i^-\over \alpha_i}=\exp\left\{{\displaystyle
\left(a_i-{a_0\over
a_i}\right)\left[x-\left(a_i+{a_0\over
a_i}\right)t-x_{0i}\right]}\right\},
\ee
\be
A_{12}=a_0\left({a_2-a_1\over a_1a_2-a_0}\right)^2.
\ee

\section{The Kaup system. Miura transformation to NLBq}
\setcounter{equation}{0}

In the previous section, we have seen how the Kaup system (KS) arises as
the Singular Manifold
equation for NLBq. That suggests a Miura transformation between KS and NLBq
\cite{estevez:LW}, \cite{estevez:kw}. Let us write the Kaup system in the form
\be
u_t=\eta_{xx}+2uu_x,
\ee
\be
\eta_t=u_{xx}+2u\eta_x.
\ee

Note that if we set $u=p_x$, (5.1-2) can be expressed as follows
\be
p_{tt}-p_{ x x x x}-4p_{ x}p_{ x t}+6p_{ x}^2p_{ x
x}-2p_tp_{ x x}=0.
\ee
This is precisely the Singular Manifold equation (4.13) for NLBq.

If we use Painlev\'e series for $u$ and $\eta$
\[
u=\sum_{j=0}^{\infty}u_j\chi^{j-a},
\]
\[
\eta=\sum_{j=0}^{\infty}\eta_j\chi^{j-b}
\]
the dominant terms yield:
$a=b=1$ and
\be
u_0=\pm \chi_x,
\ee
\be
\eta_0=\chi_x.
\ee
The existence of two branches in the Painlev\'e expansion suggests the
following change of functions:
\be
u=m-\hat m,
\ee
\be
\eta=m+\hat m.
\ee
With this change the addition and subtraction of (5.1) and (5.2) yields
\be
m_t=m_{xx}+2(m-\hat m)m_x,
\ee
\be
\hat m_t=-\hat m_{xx}+2(m-\hat m)\hat m_x.
\ee

\subsection{Miura transformation}

\noindent
From (5.8) we can obtain $\hat m$. By substituting it in (5.9), the result is:
\be
m_x^2(m_{tt}-m_{xxxx
})=4m_x^3m_{xx}+2m_x(m_tm_{xx}-m_{xx}m_{xxx})-m_{xx}(m_t^2-m_{xx}^2)
\ee
that is precisely the (4.3) NLBq.
In a similar way $m$ can be obtained from (5.9). Its substitution in (5.8)
takes the form
\be
\hat m_x^2(\hat m_{tt}-\hat m_{xxxx
})=4\hat m_x^3\hat m_{xx}+2\hat m_x(\hat m_t\hat m_{xx}-\hat
m_{xx}\hat m_{xxx})-\hat m_{xx}(\hat m_t^2-\hat m_{xx}^2)
\ee
that is again NLBq. In consequence, the splitting (5.6-7) leads to the
possibility of constructing Soliton Solutions of KS by linear superposition of
two solutions of NLBq. Actually  the inversion of (5.6-7) can be written as:
\be
2m_x=u_x-u^2+\partial_x^{-1}u_t,
\ee
\be
2\hat m_x=-u_x-u^2+\partial_x^{-1}u_t
\ee
that is a  Miura transformation between KS and NLBq \cite{estevez:LW}.

\subsection{B\"acklund  transformations}

Although two solutions $m$ and $\hat m$ of NLBq can be used to construct
(by means of (5.6-7)) a
solution of KS, these solutions are certainly not unrelated. Actually
(5.8-9) establishes the
correspondent relation between $m$ and $\hat m$. This relationship
can be written in the form:
\be
m=\hat m+{\hat m_t+\hat m_{xx}\over 2\hat m_x},
\ee
\be
\hat m=m+{m_{xx}-m_t\over 2 m_x}
\ee
which obviously is the B\"acklund  transformation that relates the two
solution of NLBq.

\subsection{Two Singular Manifolds}

The Singular Manifold approach derived in the previous section can be
applied to $m$ and $\hat m$. The Painlev\'e expansion takes the form
\be
m'=m+{\phi_x\over \phi}\Longrightarrow u'=u+{\phi_x\over \phi}-{\hat
\phi_x\over \hat \phi},
\ee
\be
\hat m'=\hat m+{\hat \phi_x\over \hat \phi}\Longrightarrow
\eta'=\eta+{\phi_x\over
\phi}+{\hat \phi_x\over \hat \phi}.
\ee
Noneless the B\"acklund transformations (5.14-15) imply that $\phi$ and
$\hat \phi$ are not unrelated. The substitution of the Painlev\'e
expansions (5.16-17) (or alternatively in (5.1-2))
gives rise to (see Appendix E) the coupling condition
\be
{\phi_x\over \phi}
{\hat \phi_x\over \hat \phi}=A{\phi_x\over \phi}
+\hat A{\hat \phi_x\over \hat \phi},
\ee
where
\be
A={v-w\over 2}+u,
\ee
\be
\hat A={\hat v+\hat w \over 2}-u,
\ee
and
\be
\lambda=\hat \lambda,
\ee
\be
u=\lambda +{\hat v+\hat w-v+w\over 2}.
\ee
By using the def\/initions (4.16-17), the expressions (5.19-22) are:
\be
A= u+\lambda+{\psi_x^-\over \psi^-},
\ee
\be
\hat A= -(u+\lambda)+{\hat \psi_x^+\over \hat \psi^+},
\ee
\be
u+\lambda={\hat \psi_x^+\over \hat \psi^+}-{\psi_x^-\over \psi^-}.
\ee

\subsection{Lax pair for KS}

The derivative of the coupling condition (5.18) with respect to $x$ is:
\be
A_x=A(\hat v-A-\hat A)= {\hat v-\hat w-v+w\over 2}={\hat \psi_x^-\over \hat
\psi^-}-{\psi_x^-\over \psi^-},
\ee
\be
\hat A_x=\hat A( v-A-\hat A)= {v+w-\hat v-\hat w\over 2}={\psi_x^+\over
\psi^+}-{\hat \psi_x^+\over \hat \psi^+}.
\ee
These expressions can be easily integrated as:
\be
A=a{\hat \psi^-\over \psi^-},
\ee
\be
\hat  A=b{ \psi^+\over \hat \psi^+}.
\ee
Combined with (5.23-24) these formulae yield
\be
\psi_x^-=a\hat \psi^- -(u+\lambda)\psi^-,
\ee
\be
\hat \psi_x^+=\hat a \psi^+ +(u+\lambda)\hat \psi^+.
\ee
The substitution of (5.25) in (5.30-31) leads to
\be
a\hat \psi^-_x=(u_x-\eta_x)\psi^-,
\ee
\be
\hat a \psi^+_x=-(u_x+\eta_x)\hat \psi^+,
\ee
where we have used $\hat \psi^+_x\hat \psi^-_x=-\hat m_x\hat
\psi^+\hat \psi^-$, $\psi^+_x \psi^-_x=- m_x \psi^+ \psi^-$ (see
Appendix D).

Those expressions can be  written as
\be
\left(\begin{array}{c}\psi^-\\ \hat \psi^-\end{array}
\right)_x
=
\left(
\begin{array}{cc} -(u+\lambda) &a\\[2mm] \displaystyle{{u_x-\eta_x\over a}}  &0
\end{array}
\right)\left(\begin{array}{c}\psi^-\\ \hat \psi^-\end{array}
\right),
\ee
 \be
\left(\begin{array}{c}\psi^+\\ \hat \psi^+\end{array}
\right)_x
=
\left(
\begin{array}{cc} 0 &\displaystyle{{-(u_x+\eta_x)}\over \hat a}\\[2mm] \hat a
&(u+\lambda)
\end{array}
\right)\left(\begin{array}{c}\psi^+\\ \hat \psi^+\end{array}
\right).
\ee
(5.34-35) are the spatial part of two components Lax pair for KS.
The temporal part can
be obtained from (4.17) and (4.19).
\be
\left(\begin{array}{c}\psi^-\\ \hat \psi^-\end{array}
\right)_t
=
\left(
\begin{array}{cc} {\displaystyle\left({1\over
2a}\right)}[\eta_{xx}-u_{xx}-(u-\lambda)(\eta_x-u_x)],
&\displaystyle{{u_x-\eta_x\over
2}}\\[2mm] -\left(\displaystyle{{\eta_x+u_x\over 2}}+u^2-\lambda^2\right),
&a(u-\lambda)
\end{array}
\right)\left(\begin{array}{c}\psi^-\\ \hat \psi^-\end{array}
\right),
\ee
 \be
\left(\begin{array}{c}\psi^+\\ \hat \psi^+\end{array}
\right)_t
=
\left(
\begin{array}{cc} \displaystyle{{u_x+\eta_x\over
2}}, &{\displaystyle\left({1\over
2\hat a}\right)}[-\eta_{xx}-u_{xx}-(u-\lambda)(\eta_x+u_x)]\\[2mm] \hat
a(u-\lambda),
&\left(\displaystyle{{\eta_x-u_x\over 2}}+u^2-\lambda^2\right)
\end{array}
\right)\left(\begin{array}{c}\psi^+\\ \hat \psi^+\end{array}
\right).
\ee

\section{Conclusions}

This paper has been dealing all along with the Painlev\'e analysis in the
version formulated by Weiss, Tabor and Carnevale \cite{estevez:WTC}. Even though
there is no rigorous proof so far available of the connection between
the Painlev\'e property and integrability  the work
hereby reviewed aims to contribute to a better understanding of the
validity and usefulness of methods based on the Painlev\'e property
for studying Nonlinear Partial Dif\/ferential Equations. With
this idea in mind  we would like to underscore some of the results that we
have obtained in this paper by applying the Singular Manifold Method
of Weiss \cite{estevez:W1}, \cite{estevez:W2}.

\begin{itemize}
\item  In Section 2, we have applied the SMM to AKNS. This
method have been proved to be quite useful to construct the Lax pair
of AKNS. By applying the SMM to the Lax pair itself Darboux
transformations and Hirota functions have been constructed
algorithmically. The use of the SMM to construct solutions
iteratively has been shown with the help of examples.

\item  A similar procedure has been used in Section 3 to study NLBq.

\item The identif\/ication of Miura transformations and B\"acklund
transformations by means of the SMM for equations with two Painlev\'e
branches appears as the main goal of Sections 3 and 5. There
sinh-Gordon and Kaup systems are presented as the modif\/ied versions
of AKNS and NLBq respectively. Two component-Lax pairs for both
systems are obtained from the AKNS and NLBq Lax pairs as induced by
the Miura map.
\end{itemize}

\newpage

\null
\newpage

\null

\newpage

\section*{Acknowledgements}
We would like to thank  Professor Jose M. Cerver\'o  for enlightening
discussions and a careful
reading of the manuscript. We thank also Professor P. Clarkson for
stimulating discussions and Dr. A. Pickering that provided us useful
references.

This research has been supported in part by DGICYT under project PB95-0947.

\appendix

\renewcommand{\theequation}{A.\arabic{equation}}
\setcounter{equation}{0}

\section{Appendix}

By substitution of (2.3) in (2.1) we obtain a polynomial in
$\ds \left({\phi_x\over\phi}\right)$
whose coef\/f\/icients are (we have used the code MAPLE V for the algebraic
computer algebra):

\medskip

\noindent  $\bullet$ Coef\/f\/icient in $\ds \left({\phi_x\over\phi}\right)^3$
\be
 4M_y+2v_y+8qM_x+qv^2+2qv_x=0.
\ee

\noindent$\bullet$ Coef\/f\/icient in $\ds \left({\phi_x\over\phi}\right)^2$
\be
\ba{l}
\ds  -4M_{xy}-2qM_{xx}-6vM_y-12qvM_x-8q_xM_x-2q_xv_x\\[2mm]
\qquad \ds -2q_xv^2-{7\over 2}vv_xq-{1\over
2}qv_{xx}-3vv_y-{3\over 2}v^3=0.
\ea
\ee

\noindent$\bullet$ Coef\/f\/icient in $\ds \left({\phi_x\over\phi}\right)$
\be
\ba{l}
\ds
8vM_{xy}+4M_{xx}(q_x+qv)+4M_y(v_x+v^2)+8M_x(qv^2+vq_x+v_y)+v_{yxx}\\[2mm]
+3vv_{xy}+ v_{xx}(q_x+qv)+3v_y(v_x+v^2)+3qv^2v_x+3vv_xq_x+v^3q_x+qv^4=0.
\ea
\ee

\noindent$\bullet$ Coef\/f\/icient in $\ds \left({\phi_x\over\phi}\right)^0$
\be
 0=M_{yxxx}+4M_yM_{xx}+8M_xM_{xy}.
\ee
(A.4) means that M is a solution of AKNS and (A.1) can be used to obtain
\be
M_y=-2qM_x-{1\over 2}v_y-{1\over 4}qv^2-{1\over 2}qv_x.
\ee
The substitution of (A.5) in (A.2) is
\[
M_{xx}+{v_{xx}\over 4}+{vv_x\over 4}=0
\]
that can be integrated as
\be
M_x=-{v_x\over 4}-{v^2\over 8}-{\lambda(t)\over 2},
\ee
where $\lambda$ is a constant for the integration with respect to $x$. The
substitution of (A.6) in (A.5) is:
\be
M_y=-{v_y\over 2}+\lambda(t)q.
\ee
 The cross derivatives of (A.6)  and (A.7) yield
\[
v_{xy}-vv_y=4\lambda q_x,
\]
\[
{d\lambda\over dt}=0
\]
that are the Singular Manifold equations.

\renewcommand{\theequation}{B.\arabic{equation}}
\setcounter{equation}{0}

\section{Appendix}

\noindent $\bullet$ Let us substitute (3.22) in (3.3). The result is:
\be
{\phi_x\over \phi}(v+2u)+{\hat\phi_x\over \hat\phi}(\hat v-2u)-2{\phi_x\over
\phi}{\hat\phi_x\over \hat\phi}=0
\ee
that compared with (3.24) reads
\be
A={v\over 2}+u,
\ee
\be
\hat A={\hat v\over 2}-u.
\ee

\noindent $\bullet$ Taking the derivative of (3.24) with respect to $x$.

\renewcommand{\theequation}{C.\arabic{equation}}
\setcounter{equation}{0}

\section{Appendix}
The substitution of (4.4) in (4.2) leads to a polynomial in
$\ds {\phi_x\over\phi}$  whose
coef\/f\/icients are:

\medskip

\noindent$\bullet$ Coef\/f\/icient in $\ds \left({\phi_x\over\phi}\right)^3$
\be
4M_{xx}-w_t+v_{xx}+ww_x+vv_x=0.
\ee

\noindent$\bullet$ Coef\/f\/icient in $\ds \left({\phi_x\over\phi}\right)^2$
\be
\ba{l}
N_t-M_{xxx}+6vM_{xx}-2M_tw-6M_x^2+M_x(w^2-v^2-4v_x)\\[2mm]
\qquad +w_x^2-v_x^2+vv_{xx}+v^2v_x+vww_x-vw_t=0.
\ea
\ee

\noindent$\bullet$ Coef\/f\/icient in $\ds \left({\phi_x\over\phi}\right)$
\be
\ba{l}
vN_t-vM_{xxx}+2M_{xx}(v_x+v^2)-2M_t(w_x+wv)-6vM_x^2\\[2mm]
\qquad +M_x(-v_{xx}-v^3+w_t-3vv_x+ww_x+w^2v)=0.
\ea
\ee

If we set $\ds w_t=\left(v_x-{v^2\over 2}+p\right)_x$, equation (C.1)
can be integrated in $x$ as:
\be
M_x={1\over 4}\left(p-v^2-{w^2\over 2}+2\lambda^2(t)\right),
\ee
where $\lambda$ is an integration constant. By multiplying (C.3) for $v$
and subtracting (C.2)
\be
2w_x(-2M_t-vww_x+2wM_x)+(v_x+2M_x)(4M_{xx}-2vv_x)=0.
\ee
The substitution of (C.4) in (C.5) yields
\be
M_t={1\over 8}\left[-4vw_x+2(z-w)v_x+(z+w)\left(p-v^2-{w^2\over
2}+2\lambda^2(t)\right)\right],
\ee
where we have equated
\be
 p_x=zw_x.
\ee

The compatibility $M_{xt}=M_{tx}$ between (C.4) and (C.7) implies that
\be
z=3w+4\lambda,
\ee
\be
p={3\over 2}w^2+4\lambda w+2\lambda^2,
\ee
\[
{d\lambda\over dt}=0,
\]
and therefore we obtain
\be
w_t=\left(v_x-{v^2\over 2}+{3\over 2}w^2+4\lambda
w+2\lambda^2\right)_x,
\ee
\be
M_x={1\over 4}\left[(w+2\lambda)^2-v^2\right],
\ee
\be
M_t={1\over 2}\left\{(w+2\lambda)v_x-
vw_x+(w+\lambda)[(w+2\lambda)^2-v^2]\right\}.
\ee

\renewcommand{\theequation}{D.\arabic{equation}}
\setcounter{equation}{0}

\section{Appendix}
To simplify the calculation let us def\/ine (see (4.16-17))
\be
2\alpha^+=2{\psi^+_x\over \psi^+}=v+w+2\lambda,
\ee
\be
2\alpha^-=2{\psi^-_x\over \psi^-}=v-w-2\lambda,
\ee
or
\be
v=\alpha^++\alpha^-,
\ee
\be
w+\lambda=\alpha^+-\alpha^-.
\ee
The substitution of (D.3-4) in (4.5-6) is
\be
M_x=-\alpha^+\alpha^-,
\ee
\be
M_t=-\alpha^+\alpha^-\left(2\alpha^+-2\alpha^-
-2\lambda+{\alpha^+_x\over \alpha^+}-
{\alpha^-_x\over \alpha^-}\right).
\ee

\noindent$\bullet$ In order to remove $\alpha^-$ from (D.5) and (D.6), we
use (D.5) to set
\be
\alpha^-=-{M_x\over \alpha^+}.
\ee
Its substitution in (D.6) provides:
\be
\ba{l}
\ds M_t=M_x\left[-{M_{xx}\over
M_x}+2{M_x\over\alpha^+}+2\alpha^++2{\alpha^+_x\over
\alpha^+}-2\lambda\right]\\[4mm]
\ds \qquad =M_x\left[-{M_{xx}\over
M_x}+2M_x{\psi^+\over\psi^+_x}+2{\psi^+_{xx}\over
\psi^+_x}-2\lambda\right]
\ea
\ee
that is (4.18). (4.20) can be obtained in the same form by removing
$\alpha^+$ between (D.5) and
(D.6).

\medskip

\noindent$\bullet$ The temporal part of the Lax pair is obtained from the
derivation of (D.1)
with respect to $t$
\be
2\alpha^+_t=2\left({\psi^+_x\over \psi^+}\right)_t=v_t+w_t.
\ee
The use of (4.9-10) yields:
\be
2\alpha^+_t=\left[w_x+wv+v_x-{v^2\over 2}+{3\over
2}(w+2\lambda)^2-2\lambda(w+2\lambda)\right]_x
\ee
that with the use of (D.3) and (D.4) f\/inally leads to
\be
\alpha^+_t=\left[\alpha^+_x+\alpha^{+2}-2\lambda
\alpha^+-4\alpha^+\alpha^-\right]_x.
\ee
Removing $\alpha^-$ with the aid of (D.7)
\be
\alpha^+_t=\left[\alpha^+_x+\alpha^{+2}-2\lambda
\alpha^+-4\alpha^+2M_x\right]_x.
\ee
Finally we can substitute (D.1) and integrate out in $x$ as
\be
{\psi^+_t\over \psi^+}={\psi^+_{xx}\over \psi^+}-2\lambda {\psi^+_x\over
\psi^+}+2M_x
\ee
that is precisely (4.19). The expression (4.21)
can be obtained by repeating the same process with (D.2).

\renewcommand{\theequation}{E.\arabic{equation}}
\setcounter{equation}{0}

\section{Appendix}

\noindent $\bullet$  The substitution of (5.16) in (5.1) gives
\be
{\phi_x\over \phi}(-w+v+2u)+{\hat\phi_x\over \hat\phi}(\hat w+\hat
v-2u)-2{\phi_x\over
\phi}{\hat\phi_x\over \hat\phi}.
\ee
The comparison with (5.18) yields to
\be
A=u+{v-w\over 2},
\ee
\be
\hat A=-u+{\hat v+\hat w\over 2}.
\ee

\noindent$\bullet$  The derivative of (5.18) with respect to $x$ provides
just like in Appendix B
\be
A_x=A(\hat v-A-\hat A),
\ee
\be \hat A_x=\hat A( v-A-\hat A).
\ee

\noindent$\bullet$  The substitution of (5.16) in (5.2) gives us:
\be
\ba{l}
\ds {\phi_x\over \phi}[-w_x-wv+v_x+v^2+2uv+2\eta_x]+{\hat \phi_x\over
\hat\phi}[-\hat w_x-\hat w\hat v-\hat v_x-\hat v^2+2u\hat v-2\eta_x]\\[3mm]
\ds \qquad +\left({\phi_x\over \phi}\right)^2[w-v-2u+2A]+
\left({\hat \phi_x\over  \hat\phi}\right)^2[\hat w+\hat v-2u-2\hat A]\\[3mm]
\ds \qquad +2{\phi_x\over \phi}{\hat \phi_x\over
\hat\phi}\left[\hat v-v+{\phi_x\over \phi}-{\hat \phi_x\over
\hat\phi}\right]=0.
\ea
\ee
The use of (E.4-5) and (5.18) yields to:
\[
\ba{l}
\ds {\phi_x\over \phi}[-w_x-wv+v_x+v^2+2uv+2\eta_x+2A(\hat v-v+\hat
A-A)]\\[3mm]
\ds \qquad +{\hat \phi_x\over
\hat\phi}[-\hat w_x-\hat w\hat v-\hat v_x-\hat v^2+2u\hat v-2\eta_x+2\hat
A(\hat v-v+\hat
A-A)]=0.
\ea
\]
Setting to $0$ both coef\/f\/icients and using (E.2-5)
\be
2\hat m_x=\eta_x-u_x=-2A(\hat v-A)=0,
\ee
\be
2m_x=\eta_x+u_x=-2\hat A( v-\hat A)=0.
\ee
The comparison between (E.8-9) and (4.5) leads to
\be
A={\hat v\over 2}\pm {\hat w+2\hat \lambda\over 2},
\ee
\be
\hat A={ v\over 2}\pm {w+2 \lambda\over 2}
\ee
that compared with (E.2-3) means that $A$ requires the plus sign and $\hat
A$ the minus sign and
\be
\lambda=\hat\lambda
\ee
which means that
\be
 u=A+{w-v\over 2}=-\hat A+{\hat v+\hat w\over 2}={1\over 2}[w-v+\hat w+\hat
v+2\lambda].
\ee

\label{estevez-lp}

\end{document}